\documentclass{PoS}

\title{Hadron spectrum of QCD with one quark flavor}

\ShortTitle{Hadron spectrum of QCD with one quark flavor}

\author{Federico Farchioni{}\speaker{}, Gernot M\"unster, %
        Tobias Sudmann, Ja\"ir Wuilloud\\
        Universit\"at M\"unster, Institut f\"ur Theoretische Physik,\\
        Wilhelm-Klemm-Strasse~9, D-48149~M\"unster, Germany\\
        E-mail: \email{farchion@uni-muenster.de}}

\author{Istv\'an Montvay\\
        Deutsches Elektronen-Synchrotron DESY, %
        Notkestr.~85, D-22603~Hamburg, Germany}

\author{Enno\,E. Scholz\\
        Physics Department, Brookhaven National Laboratory, %
        Upton,~NY~11973, USA}

\abstract{The latest results of an ongoing project for the lattice simulation of
  QCD with a single quark flavor are presented. 
  The Symanzik tree-level-improved Wilson action is adopted in the gauge sector and
  the (unimproved) Wilson action for the fermion. Results from new simulations
  with one step of Stout-smearing ($\rho=0.15$) in the fermion action
  are discussed.
  The one-flavor theory is simulated by a polynomial hybrid Monte Carlo
  algorithm (PHMC) at $\beta=4.0$ corresponding to $a=0.13\,$fm,
  on $16^3\cdot 32$ and $24^3\cdot 48$ lattices; the box-size is $L\simeq 2.1\,$fm 
  and $L\simeq 3.1\,$fm, respectively. 
  At the lightest simulated quark mass the (partially quenched) pion mass is $\sim 300$~MeV.
  The masses of the lightest bound states are computed, including the flavor singlet 
  scalar and  pseudoscalar mesons $\sigma_s$ and $\eta_s$, the scalar glueball $0^{++}$, 
  and the $\Delta^{++}$ baryon. 
    Relics of SUSY in the mass spectrum, expected from a large $N_c$ orientifold 
  equivalence with the ${\cal N}$=1 supersymmetric Yang-Mills theory, are discussed.}

\FullConference{The XXVI International Symposium on Lattice Field Theory\\
		 July 14-19 2008\\
		 Williamsburg, Virginia, USA}

\begin{document}

%%%%%%%%%%%%%%%%%%%%%%%%%%%%%%%%%%%%%%%%%%%%%%%%%%%%%%%%%%%%%%%%%%%%%%%%
\newcommand{\be}{\begin{equation}}
\newcommand{\ee}{\end{equation}}
\newcommand{\bea}{\begin{eqnarray}}
\newcommand{\eea}{\end{eqnarray}}
\newcommand{\half}{\frac{1}{2}}
\newcommand{\rar}{\rightarrow}
\newcommand{\lar}{\leftarrow}
\newcommand{\LCB}{\raisebox{-0.3ex}{\mbox{\LARGE$\left\{\right.$}}}
\newcommand{\RCB}{\raisebox{-0.3ex}{\mbox{\LARGE$\left.\right\}$}}}
\newcommand{\LSB}{\raisebox{-0.3ex}{\mbox{\LARGE$\left[\right.$}}}
\newcommand{\RSB}{\raisebox{-0.3ex}{\mbox{\LARGE$\left.\right]$}}}
\newcommand{\tr}{{\rm Tr}}
\newcommand{\I}{\ensuremath{\mathrm{i}\,}}
\newcommand{\E}{\ensuremath{\mathrm{e}\,}}
%%%%%%%%%%%%%%%%%%%%%%%%%%%%%%%%%%%%%%%%%%%%%%%%%%%%%%%%%%%%%%%%%%%%%%%%

\section{Introduction and motivation}

The low-energy dynamics of QCD with a single quark flavor ($N_f=1$ QCD) 
strongly differs from the one of the multi-flavor theory since
chiral symmetry, and its spontaneous breaking, is absent.
Nevertheless, several open questions of the physical theory may be 
better understood in the theory with minimal number of fermionic degrees of freedom
\cite{Creutz:nf1,Creutz:UpMass,Creutz:CP}. 
For example \cite{Creutz:UpMass}, whether setting to zero just one quark mass
produces physical effects;
in the one-flavor model, due to the lack of a chiral symmetry, even the definition
of the quark mass is non-trivial \cite{Creutz:nf1}. Another question \cite{Creutz:CP} is 
whether spontaneous breaking of CP is possible   
for special combinations of the light quark masses
(negative quark mass for one flavor).

Another intriguing aspect of one-flavor QCD, emerging from string theory, is the connection 
with the ${\cal N}$=1 supersymmetric Yang-Mills theory (SYM),
also investigated by our collaboration~\cite{Poster}.
The equivalence of the two theories in the bosonic sector can be proved at the planar 
level of an ``orientifold'' large $N_c$ limit~\cite{ArmShiVen}.
Relics of SUSY are therefore expected in $N_f=1$ QCD (with $N_c=3$) where 
approximately degenerate scalars with opposite parity should be observed.
These two particles can be identified in the single-flavor theory with the singlet
scalar and pseudoscalar mesons which we denote with $\sigma_s$ and $\eta_s$.

As we have argued in~\cite{OurNf1}, some chiral symmetry can be recovered 
in  $N_f=1$ QCD by embedding the theory in a larger partially quenched (PQ)
theory with additional valence quark flavors. 
This allows to build pions and obtain 
a possible definition of the quark mass by the
PCAC relation.

We present here the latest results of an ongoing computation of the low-lying hadron spectrum
of $N_f=1$ QCD in the Wilson formulation~\cite{OurNf1}, 
obtained from  $16^3 \cdot 32$ and  $24^3 \cdot 48$ 
lattices at lattice spacing $a \simeq 0.13\,{\rm fm}$ ($L=2.1$ and $3.1{\rm\, fm}$). 
The lightest simulated pion mass is $M_\pi \simeq 300{\rm\,MeV}$.

\section {Partial quenching}
\label{sec:pq}

The definition of a quark mass is not immediate in one-flavor QCD
due to the lack of chiral symmetry (even in the continuum).
For example a PCAC quark mass cannot be defined. More generally, the bilinear quark
operator $\bar qq$ is not protected against scheme-dependent additive 
renormalizations and the concept of vanishing quark mass could be devoid 
of physical meaning~\cite{Creutz:nf1}. 
In our approach we take $N_V=2$ valence quarks degenerate with the
(single) sea quark. 
With the $N_V\geq 2$ valence quarks plus the sea quark, all mesons and baryons of QCD can be build,
appearing in  degenerate multiplets due to the 
exact ${\rm SU}(N_V+1)$ flavor symmetry. 
A PCAC quark mass can be defined in the PQ theory by the non-singlet axial-vector 
chiral current $A^a_{x\mu}$ 
\begin{equation}\label{eq:pcac}
m_{\rm\scriptscriptstyle PCAC} \equiv
\frac{\langle \partial^\ast_\mu A^+_{x\mu}\, P^-_y \rangle}
{2\langle P^+_x\, P^-_y \rangle} \ .
\end{equation}
Relation~(\ref{eq:pcac}) should be considered here just as a
possible definition of the quark mass in $N_f=1$ QCD (a possible physical meaning 
of this definition is not claimed at this stage).

In~\cite{OurNf1}, validity of the GMOR relation for the pions in the 
PQ extension of the one-flavor theory was confirmed: the masses of
the pions can be made to vanish by suitably tuning the bare quark mass on the lattice;
in this situation the quark mass~(\ref{eq:pcac}) vanishes, too. 
This scenario is confirmed by a study of the chiral Ward identities~\cite{BeGoShaSha}.

Although pions do not belong to the unitary sector of the theory,
their properties and the PCAC quark mass can be used for
a characterization of the one-flavor theory. In particular
the low-energy coefficients of the chiral Lagrangian
can be extracted by an analysis in PQ
chiral perturbation theory~\cite{OurNf1}. 

%%%%%%%%%%%%%%%%%%%%%%%%%%%%%%%%%%%%%%%%%%%%%%%%%%%%%%%%%%%%%%%%%%%%%%%%
\begin{table}
\begin{center}
  \caption{\label{tabruns} Summary of the runs: 
      the bar indicates runs with Stout-link in the fermion action 
      (see text).}

\vspace{1em}

\renewcommand{\arraystretch}{1.0}
\begin{tabular}{l|c*{6}{|c}}
 \multicolumn{1}{c|}{}&
 \multicolumn{1}{c|}{$L$} &
 \multicolumn{1}{c|}{$\beta$} &
 \multicolumn{1}{c|}{$\kappa$} &
 \multicolumn{1}{c|}{$N_{\textrm{\scriptsize conf}}$} &
 \multicolumn{1}{c|}{${\textrm{ plaquette}}$} &
 \multicolumn{1}{c|}{$\tau_{\textrm{\scriptsize plaq}}$} &
 \multicolumn{1}{c}{$r_0/a$}
 \\
\hline\hline
 $a$ &12& 3.80 & 0.1700 & 5424 & 0.546041(66) & 12.5 & 2.66(4)
 \\ \hline
 $b$ &12& 3.80 & 0.1705 & 3403 & 0.546881(46) & 4.6  & 2.67(5)
 \\ \hline
 $c$ &12& 3.80 & 0.1710 & 2884 & 0.547840(67) & 7.6  & 2.69(5)
 \\ \hline\hline
 $A$ &16& 4.00 & 0.1600 & 1201 & 0.581427(36) & 4.3  & 3.56(5)
 \\ \hline
 $B$ &16& 4.00 & 0.1610 & 1035 & 0.582273(36) & 4.1  & 3.61(5)
 \\ \hline
 $C$ &16& 4.00 & 0.1615 & 1005 & 0.582781(32) & 3.3  & 3.73(5)
 \\ \hline\hline
 $\bar A$ &16& 4.00 & 0.1440 & 5600 & 0.577978(23) & 9.7  & 3.74(3)
 \\ \hline
 $\bar B$ &16& 4.00 & 0.1443 & 5700 & 0.578167(28) & 11.3  & 3.83(5) 
 \\ \hline
 $\bar B_{24}$ &24& 4.00 & 0.1443 & 3900 & 0.578182(10)  & 5.8 & 3.83(4)
 \\ \hline\hline
\end{tabular}

\vspace{-1em}

\end{center}
\end{table}
%%%%%%%%%%%%%%%%%%%%%%%%%%%%%%%%%%%%%%%%%%%%%%%%%%%%%%%%%%%%%%%%%%%%%%%%

\section{Simulation}
\label{sec:sim}

The gauge action is discretized by the tree-level improved Symanzik (tlSym) lattice 
action~\cite{WeiszWohlert} including planar rectangular $(1\times 2)$ Wilson loops.
We apply the (unimproved) Wilson formulation in the fermionic sector.
Previous results~\cite{OurNf1} were presented for simulations with the original Wilson 
formulation; with the goal of improving stability of the Monte Carlo evolution at small quark masses, 
we now apply Stout-smeared links~\cite{Stout} in the hopping matrix
(one step of isotropic smearing with 
coefficient $\rho_{\mu\nu}=\rho=0.15$).

The update algorithm is a Polynomial Hybrid Monte Carlo algorithm 
with a two-step polynomial approximation (TS-PHMC) \cite{MontvayScholz}.  
A correction factor $C[U]$ in the measurement is associated to 
configurations for which  the eigenvalues of the (squared Hermitian) fermion matrix 
lie outside the validity range of the polynomial approximation.
See~\cite{OurNf1} for more details on the algorithmic setup.
The sign of the determinant associated to one (light) Wilson 
quark  can become negative on some configurations, even for positive quark masses. 
Since the sign cannot be taken into account at the update level, it must be computed ``off-line''
and included in the correction factor; the expectation
value of a generic quantity $A$ is therefore given by
\be\label{eq08}
\langle A \rangle = \frac{\int [dU]\:\sigma[U]\,C[U]\,A[U]}
                         {\int [dU]\:\sigma[U]\,C[U]}  \ ,
\ee
where $\sigma[U]$ is the sign of the one-flavor determinant. 
For the computation of the sign we study the 
(complex) spectrum of the non-Hermitian matrix concentrating on the lowest real eigenvalues:
sign changes are signaled by negative real eigenvalues. We applied
the ARPACK Arnoldi routines on a transformed Dirac operator.
The (polynomial) transformation was tuned such that the real eigenvalues are
projected outside the ellipsoidal bulk containing the whole eigenvalue spectrum~\cite{Saad}.
This allows an efficient computation of the real eigenvalues~\cite{Neff}.

%%%%%%%%%%%%%%%%%%%%%%%%%%%%%%%%%%%%%%%%%%%%%%%%%%%%%%%%%%%%%%%%%%%%%%%%
\begin{table}[t]
\begin{center}
\caption{\label{tab:hadmass}
 Results for hadron observables in $N_f=1$ QCD (lattice units).}

%\vspace{1em}

\renewcommand{\arraystretch}{1.0}

\begin{tabular}{l||l|l|l||l|l|l|l}
 \multicolumn{1}{c||}{}&
 \multicolumn{1}{c|}{$aM_{\eta_s}$} &
 \multicolumn{1}{c|}{$aM_{\sigma_s}$} &
 \multicolumn{1}{c||}{$aM_{\Delta_s}$} &
 \multicolumn{1}{c|}{$am_{\rm\scriptscriptstyle PCAC}$} &
 \multicolumn{1}{c|}{$aM_\pi$} &
 \multicolumn{1}{c|}{$af_\pi$} &
 \multicolumn{1}{c}{$aM_N$} 
 \\
\hline\hline
 $a$               & 0.462(13)  & 0.660(39)  &  1.215(20)  & 0.0277(5)    & 0.3908(24) & 0.1838(11)  & 1.044(5)
 \\ \hline
 $b$               & 0.403(11)  & 0.629(29)  &  1.116(38)  & 0.0195(4)    & 0.3292(25) & 0.1730(15)  & 0.956(3)%(27)
 \\ \hline 
 $c$               & 0.398(28)  & 0.584(55)  &  1.204(57)  & 0.0108(12)   & 0.253(10)  & 0.156(10)   & 1.01(5)
 \\ \hline\hline
 $A$               & 0.455(17)  & 0.607(57)  &  1.006(15)  & 0.04290(36)  & 0.4132(21)  & 0.1449(9)  & 0.902(4)
 \\ \hline
 $B$               & 0.380(18)  & 0.554(52)  &  0.960(15)  & 0.02561(31)  & 0.3199(22)  & 0.1289(10) & 0.798(5)
 \\ \hline
 $C$               & 0.344(21)  & 0.576(53)  &  0.971(30)  & 0.01681(33)  & 0.2622(19)  & 0.1190(17) & 0.762(9)
 \\ \hline\hline
 $\bar A$        & 0.347(16)  & 0.538(41)  &  0.855(50)  & 0.01651(27)  & 0.2471(19)  & 0.0983(20) & 0.733(13)
 \\ \hline
 $\bar B$        & 0.286(18)  & 0.485(46)  &  0.848(70)  & 0.01094(23)  & 0.2028(35)  & 0.0913(24) & 0.670(20) 
 \\ \hline
 $\bar B_{24}\!\!$  & 0.261(11) & 0.496(22)&    0.900(24) & 0.01047(17)  & 0.1958(15)   & 0.0920(11) &  0.672(10)        
 \\ \hline\hline
\end{tabular}

\vspace{-1em}

\end{center}
\end{table}
%%%%%%%%%%%%%%%%%%%%%%%%%%%%%%%%%%%%%%%%%%%%%%%%%%%%%%%%%%%%%%%%%%%%%%%%

%\subsection{Simulation details}

The updating has been performed by applying determinant break-up
by a factor of two  (that is, two half-flavors were considered
instead of one).
The update sequence consisted of two PHMC trajectories followed
by an accept-reject step by the second (precise) polynomial
approximation.
The precision of the first polynomial approximation was tuned
such that an acceptance of about 90\% was obtained.
The same acceptance was required for the Metropolis test at the
end of every individual PHMC trajectory by tuning the trajectory
length (0.4-0.5).
This resulted in a high total acceptance rate of about 80\%.
The gauge configurations were stored after every accept-reject step
by the precise polynomial.
Relatively short trajectories were chosen in order to have many
configurations for the glueball mass determination.
Optimization of the parameters of PHMC turned out to have a substantial 
impact on the integrated autocorrelation times of the average plaquette.
Information about the sets of configurations generated up to now can be found in 
Table~\ref{tabruns}. New results presented in this Contribution 
concern runs $C,\bar A,\bar B, \bar B_{24}$.

We fix the lattice scale by the Sommer scale parameter $r_0$~\cite{Sommer}. 
For the conversion into physical units we assume the conventional value $r_0=0.5{\rm\: fm}$.
As in ordinary QCD, $r_0/a$ grows with the hopping parameter $\kappa$; 
consistently with a massless scheme the value of
$r_0/a$ should be extrapolated to the critical value $\kappa=\kappa_c$
where the PCAC quark mass (\ref{eq:pcac}) vanishes. 
The number of quark masses at our disposal are however not sufficient for an 
extrapolation and we rely on the value at the highest $\kappa$ for
given $\beta$.
We obtain in this way $a(3.8) \simeq 0.19{\rm\, fm}$ and $a(4.0) \simeq 0.13{\rm\: fm}$.
This corresponds to a roughly constant volume $L=2.1{\rm\, fm}$.
In order to check finite volume effects, a run on a larger $24^3\cdot 48$ lattice 
has been started, see Table~\ref{tabruns}, corresponding to $L=3.1{\rm\, fm}$. 

We observe that, for a fixed quark mass (in lattice units), 
the Stout-smearing leaves $r_0/a$ essentially unchanged
(as can be seen by comparing runs $C$ and $\bar A$).
On the basis of this observation, we will assume 
approximate equality of the lattice spacing
for the  $\beta=4.0$ runs with and without smearing; therefore data from both sets will be included 
in analyzes at fixed lattice spacing.

\section{Hadron spectrum}
\label{sec:spec}

The disconnected diagrams of the $\eta_s$ and $\sigma_s$ correlators 
(with usual interpolating fields) were computed by 
applying for each configuration 20 stochastic sources 
with complex $Z_2$ noise and spin dilution. 
In consideration of autocorrelations, we analyzed every 10th, 16th or 32th configuration 
according to $\kappa$. The resulting statistics vary
between 400 and 600.

In the case of the baryon, we take the interpolating field
$
 {\Delta_i}(x)\:=\:\epsilon_{abc}[\psi_a(x)^TC\gamma_i\psi_b(x)]\psi_c(x)\ .
$
The  low-lying projected state is expected to be a spin $3/2$ parity-positive particle
which we denote with $\Delta_s$ (the desired spin 3/2 component is extracted by spin-projection).
It corresponds to the $\Delta^{++}(1232)$ baryon of QCD if the single quark is identified with the 
$u$-quark.

%%%%%%%%%%%%%%%%%%%%%%%%%%%%%%%%%%%%%%%%%%%%%%%%%%%%%%%%%%%%%%%%%%%%%%%
\begin{figure}[t]
  %\centering
\begin{center}
 \hspace{0cm} \includegraphics[angle=0,width=.73\linewidth]{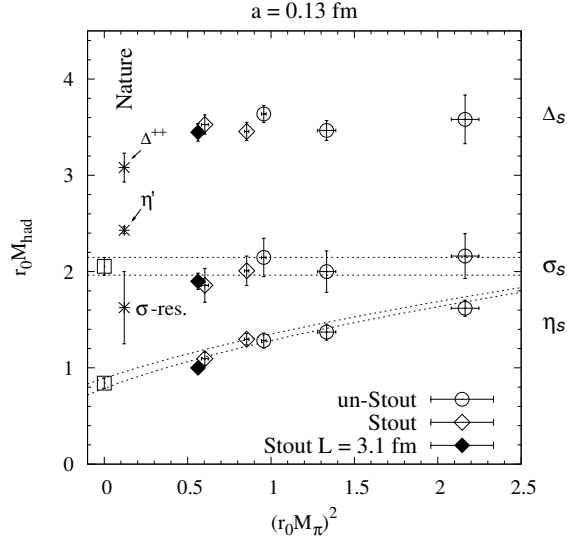}
\vspace*{-1em}

  \caption{\label{fig:hadmass}%\em
    The mass of the lightest physical particles in one-flavor QCD as
    a function of the squared pion mass in units of the Sommer scale.  
   The hadron masses are multiplied by the value of $r_0/a$ at the given $\kappa$.
}

\vspace*{-1em}

\end{center}
\end{figure}
%%%%%%%%%%%%%%%%%%%%%%%%%%%%%%%%%%%%%%%%%%%%%%%%%%%%%%%%%%%%%%%%%%%%%%%

In the PQ sector we measure the pion observables, 
$M_\pi, f_\pi, m_{\rm\scriptscriptstyle PCAC}$, in which case 
correlators trivially coincide with the connected parts of the 
corresponding $\eta_s$ correlators; the nucleon mass 
is determined by applying the standard projecting operator~\cite{OurNf1}. 

No smearing was applied for the extraction of the masses.
It turns out however that this is necessary in the baryon sector where the 
approach to the asymptotic behavior is slow. 
The optimization of the overlap with the ground state by Jacobi smearing  
is in plan.

%%%%%%%%%%%%%%%%%%%%%%%%%%%%%%%%%%%%%%%%%%%%%%%%%%%%%%%%%%%%%%%%%%%%%%%
\begin{figure}[t]
  %\centering
\begin{center}
   \hspace{0cm} \includegraphics[angle=0,width=.73\linewidth]{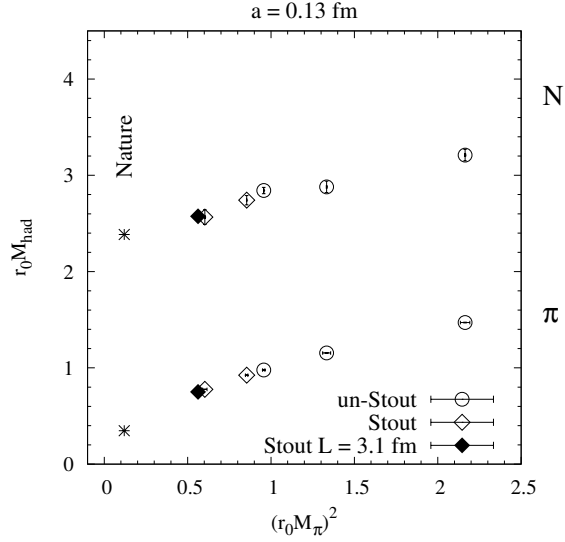}
  \vspace*{-1em}

  \caption{\label{fig:pqhadmass}%\em
    The PQ sector of one-flavor QCD: the nucleon mass as
    a function of the square pion mass  in units of the Sommer scale.  
    The pion mass is also reported for comparison.}

\vspace*{-1em}

\end{center}
\end{figure}
%%%%%%%%%%%%%%%%%%%%%%%%%%%%%%%%%%%%%%%%%%%%%%%%%%%%%%%%%%%%%%%%%%%%%%%

\vspace{1mm}

{\bf Results.}\ \ 
The results for the hadron observables in lattice units including new  
runs with Stout-smearing are reported in Table~\ref{tab:hadmass}. 
The lightest pion mass corresponds to $\sim {\rm \, 300 MeV}$.
From the comparison between runs $\bar B$ and $\bar B_{24}$
 ($L=2.1{\rm\, fm}$ and $L=3.1{\rm\, fm}$, respectively) 
finite volume effects can be estimated. 
These are below the statistical accuracy ($\sim 10\%$)
for the particles in the unitary sector. 
In the pion sector, they exceed statistical accuracy 
in the case of the pion mass and of the PCAC quark mass, however 
they are relatively small ($\sim4\%$ in both cases).
Observe that the sign 
of the finite size scaling agrees with what is observed in standard QCD.

In Fig.~\ref{fig:hadmass} the hadron masses in the unitary sector
are reported in physical units as a function of the squared pion mass. 
The $\sigma_s$ and the $\Delta_s$ masses are (surprisingly) near to the 
values observed in nature for the corresponding particles or resonances. In contrast,
the $\eta_s$ meson is much lighter than the QCD flavor-singlet $\eta^\prime$;
this can be understood~\cite{OurNf1} in terms of the Witten-Veneziano formula.
The $\eta_s$ mass, which shows a clear quark mass dependence,
can be extrapolated to zero quark mass by applying   
the LO chiral perturbation theory formula~\cite{BeGo,OurNf1} 
$
M_{\eta_s}^2\:=\:(M_\phi^2+M_\pi^2)/(1+\alpha)\ , 
$
with $M_\phi$ and $\alpha$ constants.
The result is: $r_0M_{\eta_s}(m_q=0) = 0.84(5)\ [330(20){\rm\, MeV}]$; the error also includes
a rough estimate of the extrapolation uncertainty, obtained by comparing results 
including or excluding the heaviest quark mass. 
It is interesting to compare this result for the $\eta_s$ mass
with the one obtained by our collaboration for the corresponding particle in SU(2) SYM,
the {\em adjoint}  $\eta^\prime$ ($a\mbox{-}\eta^\prime$): 
 $r_0M_{a\mbox{-}\eta^\prime} = 1.25(5)\ [499(20){\rm\, MeV}]$~\cite{Poster}.
Since the quark mass dependence of the $\sigma_s$ mass cannot be seen with 
our statistical accuracy, we simply apply a fit to a constant; the result is:  
$r_0M_{\sigma_s} = 2.05(9)\ [810(35){\rm\, MeV}]$.

We can now check our results against the prediction from the orientifold planar 
equivalence \cite{ArmoniImeroni}: 
$M_{\eta_s}/M_{\sigma_s}=(N_c-2)/N_c\times (1+\delta)$,
where the correction $\delta=O(1/N_c,1/N_c^2)$ is expected to be suppressed.
We obtain: 
\be\label{eq:ratio}
\frac{M_{\eta_s}}{M_{\sigma_s}}=0.410(32)(25)\, ,\quad \delta=0.23(10)(7)\ .
\ee
(the second error comes from the extrapolation, only data from the $L=2.1 {\rm \, fm}$ volume
are included). 
The relatively small value of $\delta$ seems to confirm suppressed $O(1/N_c)$ 
corrections, as expected on 
theoretical grounds~\cite{ArmoniImeroni}. Small deviations from the leading orientifold prediction
were also observed for the fermion condensate, in numerical simulations~\cite{DeGrandHoffmannSchaeferLiu}, and in an
analytical computation for free staggered fermions~\cite{ArmLuPaPi}.  
The inclusion in the analysis of smaller quark masses and larger volumes could further lower
the ratio~(\ref{eq:ratio}) and therefore the deviation $\delta$.

Purely gluonic operators, the glueballs, project onto Spin~0 states, too.
We investigate here the $0^{++}$ state, which is expected to mix with the $\sigma_s$.
We neglect for the moment possible mixings with the mesonic state and consider 
diagonal correlators only.  
Since the computational load is low, we analyze in this case each configuration; this allows
to obtain a decent signal for the case of the Stout-runs, where fluctuations are
reduced. We obtain $r_0M_{0^{++}}\!=\!1.94(25)$ for run $\bar A$ and $2.43(35)$ for run $\bar B$;
these results are in the ballpark of the $\sigma_s$ mass in accordance with
strong mixing.

The behavior of the nucleon mass as a function of the pion mass squared is reported 
in Fig.~\ref{fig:pqhadmass}; for  comparison, we also report the pion mass.
Also in the case of the nucleon we observe a surprising agreement with the expectations
from the physical world.

\section{Conclusions and perspectives}
\label{sec:outlook}

New data from simulations of lighter quark masses in sufficiently large volumes 
allowed first quantitative estimates for the hadron spectrum of $N_f=1$ QCD.
Results for the $\sigma_s$ and $\eta_s$ masses could be compared
with the predictions from the orientifold equivalence; the deviation from the leading formula
for the ratio of the masses turns out to be relatively small, in accordance with 
observations for the fermion condensate~\cite{DeGrandHoffmannSchaeferLiu,ArmLuPaPi}.
With the exception of a lighter $\eta_s$, no other striking deviation from the 
physical (multi-flavor) picture is observed for the measured quantities. 
The simulation of additional lighter quark masses
in view of an analysis in chiral perturbation theory is planned for the future. 
The search for the expected CP-violating phase transition is ongoing.

\vspace{1mm}

The computations were carried out on Blue Gene L/P and JuMP systems at JSC J\"ulich (Germany).

\end{document}